# CoroNet: A deep neural network for detection and diagnosis of COVID-19 from chest x-ray images

Asif Iqbal Khan[a,*], Junaid Latief Shah[b], Mohammad Mudasir Bhat[c]

[a] *Department of Computer Science, Jamia Millia Islamia, New Delhi, India*
[b] *Higher Education Department, J&K, India*
[c] *Lelafe IT Solutions, J&K, India*

## Abstract

*Background and Objective:* The novel Coronavirus also called COVID-19 originated in Wuhan, China in December 2019 and has now spread across the world. It has so far infected around 1.8 million people and claimed approximately 114,698 lives overall. As the number of cases are rapidly increasing, most of the countries are facing shortage of testing kits and resources. The limited quantity of testing kits and increasing number of daily cases encouraged us to come up with a Deep Learning model that can aid radiologists and clinicians in detecting COVID-19 cases using chest X-rays.

*Methods:* In this study, we propose CoroNet, a Deep Convolutional Neural Network model to automatically detect COVID-19 infection from chest X-ray images. The proposed model is based on Xception architecture pre-trained on ImageNet dataset and trained end-to-end on a dataset prepared by collecting COVID-19 and other chest pneumonia X-ray images from two different publically available databases.

*Results:* CoroNet has been trained and tested on the prepared dataset and the experimental results show that our proposed model achieved an overall accuracy of 89.6%, and more importantly the precision and recall rate for COVID-19 cases are 93% and 98.2% for 4-class cases (COVID vs Pneumonia bacterial vs pneumonia viral vs normal). For 3-class classification (COVID vs Pneumonia vs normal), the proposed model produced a classification accuracy of 95%. The preliminary results of this study look promising which can be further improved as more training data becomes available.

*Conclusion:* CoroNet achieved promising results on a small prepared dataset which indicates that given more data, the proposed model can achieve better results with minimum pre-processing of data. Overall, the proposed model substantially advances the current radiology based methodology and during COVID- 19 pandemic, it can be very helpful tool for clinical practitioners and radiologists to aid them in diagnosis, quantification and follow-up of COVID-19 cases.

## INTRODUCTION

The 2019 novel Coronavirus or COVID-19, first reported in Wuhan, China in December 2019 belongs to the family of viruses "Coronavirus" (CoV) was called "Severe Acute Respiratory Syndrome Coronavirus 2" (SARS-CoV-2) before it was named COVID-19 by World Health Organization (WHO) in February 2020. The outbreak was declared a Public Health Emergency of International Concern on 30 January 2020 [1] and finally on March 11, 2020, WHO declared COVID-19 as Pandemic. After the outbreak, the number of daily cases began to increase exponentially and reached 1.8 million cases and around 114698 deaths globally by 12 April 2020. The virus has engulfed more than 210 countries among which USA, Spain and Italy are severely hit with 560,433, 166,831 and 156,363 active cases and 22,115, 17,209 and 19,899 deaths respectively [2].

Once infected, a COVID-19 patient may develop various symptoms and signs of infection which include fever, cough and respiratory illness (like flu). In severe cases, the infection may cause pneumonia, difficulty breathing, multi-organ failure and death [2,3]. Due to the rapid and increasing growth rate of the COVID-19 cases, the health system of many advanced countries has come to the point of collapse. They are now facing shortage of ventilators and testing kits. Many countries have declared total lockdown and asked its population to stay indoors and strictly avoid gatherings.

[*] Corresponding author.
   *E-mail address:* khanasifiqbal7@gmail.com (A.I. Khan).

A critical and important step in fighting COVID-19 is effective screening of infected patients, such that positive patients can be isolated and treated. Currently, the main screening method used for detecting COVID-19 is real-time reverse transcription polymerase chain reaction (rRT-PCR) [4,5]. The test is done on respiratory samples of the patient and the results can be available within few hours to 2 days. An alternate method to PCR screening method can be based on chest radiography images. Various research articles published in Radiology journal [6,7] indicate that that chest scans might be useful in detecting COVID-19. Researchers found that the lungs of patients with COVID-19 symptoms have some visual marks like ground-glass opacities—hazy darkened spots that can differentiate COVID-19 infected patients from non COVID-19 infected ones [8,9]. The researchers believe that chest radiology based system can be an effective tool in detection, quantification and follow-up of COVID-19 cases.

A chest radiology image based detection system can have many advantages over conventional method. It can be fast, analyze multiple cases simultaneously, have greater availability and more importantly, such system can be very useful in hospitals with no or limited number of testing kits and resources. Moreover, given the importance of radiography in modern health care system, radiology imaging systems are available in every hospital, thus making radiography based approach more convenient and easily available.

Today, researchers from all around the world, from various different fields are working day and night to fight this pandemic. Many researchers have published series of preprint papers demonstrating approaches for COVID-19 detection from chest radiography images [10,11]. These approaches have achieved promising results on a small dataset but by no means are production ready solutions. These approaches still need rigorous testing and improvement before putting them in use. Subsequently, a large number of researchers and data scientists are working together to build highly accurate and reliable deep learning based approaches for detection and management of COVID-19 disease. Researchers are focusing on deep learning techniques to detect any specific features from chest radiography images of COVID-19 patients. In recent past, deep learning has been very successful in various visual tasks which include medical image analysis as well. Deep learning has revolutionized automatic disease diagnosis and management by accurately analyzing, identifying, classifying patterns in medical images. The reason behind such success is that deep learning techniques do not rely on manual handcrafted features but these algorithms learn features automatically from data itself [12]. In the past, deep learning has had success in disease classification using chest radiography image. ChexNet[13] is a deep neural network model that detects Pneumonia from chest X-ray image. ChexNet achieved exceptional results exceeding average radiologist performance. Another similar approach called ChestNet [14] is a deep neural network model designed to diagnose thorax diseases on chest radiography images.

The success of AI based techniques in automatic diagnosis in the medical field and rapid rise in COVID-10 cases have necessitated the need of AI based automatic detection and diagnosis system. Recently, many researchers have used radiology images for COVD-19 detection. A deep learning model for COVID-19 detection (COVID-Net) proposed by Wang and Wong [10] obtained 83.5% accuracy in classifying COVID-19, normal, pneumonia-bacterial and pneumonia-viral classes. Hemdan et al. [15] used various deep learning models to diagnose COVID-19 from cheat X-ray images and proposed a COVIDX-Net model comprising seven CNN models. Apostolopoulos and Mpesiana [16] trained different pre-trained deep learning models on a dataset comprising of 224 confirmed COVID-19 images and achieved 98.75% and 93.48% accuracy for two and three classes, respectively. Narin et al. [11] trained ResNet50 model using chest X-ray images and achieved a 98% COVID-19 detection accuracy for two classes. However, the performance for multi class classification is not known. Sethy and Behera [17] used various convolutional neural network (CNN) models along support vector machine (SVM) classifier for COVID-19 classification. Their study states that the ResNet50 model with SVM classifier provided the best performance. Most recently Ozturk et al. [18] proposed a deep network based on DarkNet model. Their model consists of 17 convolution layers with Leaky RelU as activation function. Their model achieved an accuracy of 98.08% for binary classes and 87.02% for multi-class cases. All these techniques except COVID-Net [10] either perform binary classification (normal vs COVID-19) or 3–class classification (normal vs pneumonia vs COVID-19). Other than COVID-Net, none of the methods discussed above treat pneumonia bacterial and pneumonia viral as separate classes.

**Table I: Dataset Summary**

| Disease | No. of Images |
|---|---|
| Normal | 310 |
| Pneumonia Bacterial | 330 |
| Pneumonia Viral | 327 |
| COVID-19 | 284 |

In this study, we present a deep learning based approach to detect COVID-19 infection from chest X-ray images. We propose a deep convolutional neural network (CNN) model to classify three different types of Pneumonia; bacterial pneumonia, viral pneumonia and COVID-19 pneumonia. We also implemented binary and 3-class versions of our proposed model and compared the results with other studies in the literature. The proposed model is called CoroNet and will help us identifying the difference between three types of pneumonia infections and how COVID-19 is different from other infections. A model that can identify COVID-19 infection from chest radiography images can be very helpful to doctors in the triage, quantification and follow-up of positive cases. Even if this model does not completely replace the existing testing method, it can still be used to bring down the number of cases that need immediate testing or further review from experts.

**DATASET**

Deep learning is all about data which serves as fuel in these learning models. Since COVID-19 is a new disease, there is no appropriate sized dataset available that can be used for this study. Therefore, we had to create a dataset by collecting chest X-ray images from two different publically available image databases. COVID-19 X-ray images are available at an open source Github repository by Joseph et al [19]. The authors have compiled the radiology images from various authentic sources (Radiological Society of North America (RSNA), Radiopaedia etc) of COVID-19 cases for research purpose and most of the studies on COVID-19 use images from this source. The repository contains an open database of COVID-19 cases with chest X-ray or CT images and is being updated regularly. At the time of writing this paper, the database contained around 290 COVID-19 chest radiography images. Pneumonia bacterial, Pneumonia viral and normal chest X-ray images were obtained from Kaggle repository "Chest X-Ray Images (Pneumonia)" [20]. The dataset consists of 1203 normal, 660 bacterial Pneumonia and 931 viral Pneumonia cases. We collected a total of 1300 images from these two sources. We then resized all the images to the dimension of 224 x 224 pixels with a resolution of 72 dpi. Table I shows the summary of the prepared dataset. Figure 1 below shows some samples of chest X-ray images from the prepared dataset.

In order to overcome the unbalanced data problem, we used resampling technique called random under-sampling which involves randomly deleting examples from the majority class until the dataset becomes balanced. We used only 310 normal, 330 pneumonia-bacterial and 327 Pneumonia-viral X-ray images randomly from this chest X-ray pneumonia database.

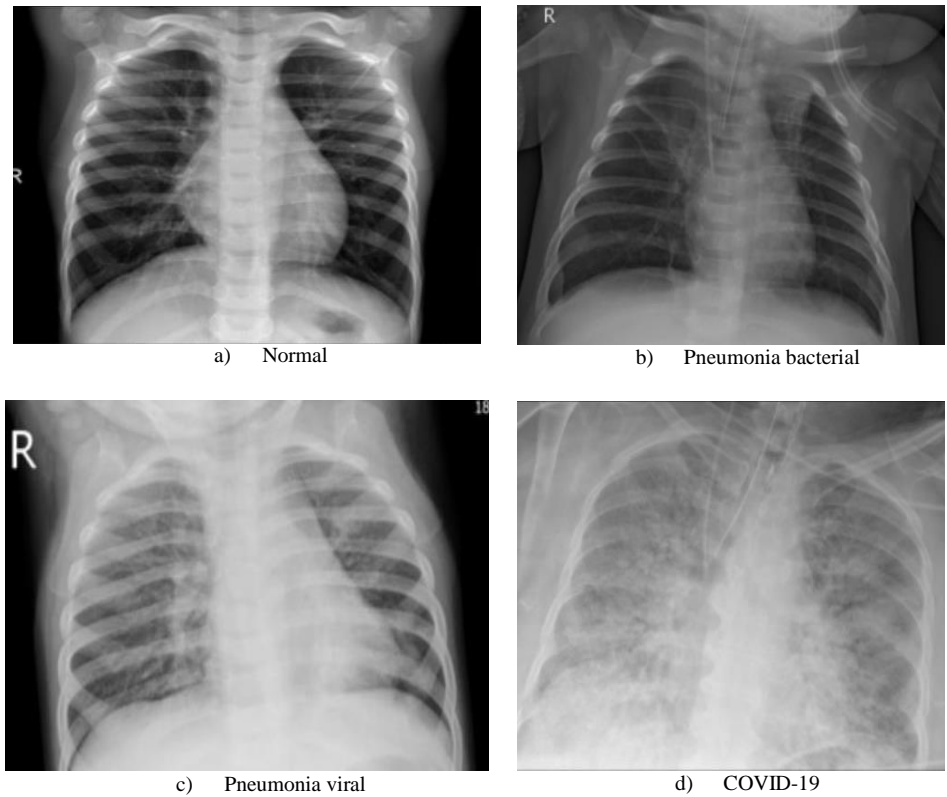

Figure 1: Samples of chest x-ray images from prepared dataset (a) Normal (b) Pneumonia bacterial (c) Pneumonia viral (d) COVID-19

**METHODOLOGY**

In this section, we will discuss the work methodology for the proposed technique, model architecture, implementation and training. The work methodology is also illustrated in Figure 2.

**Convolutional Neural Network (CNN)**

Convolutional Neural Network also known as CNN is a deep learning technique that consists of multiple layers stacked together which uses local connections known as *local receptive field* and *weight-sharing* for better performance and efficiency. The deep architecture helps these networks learn many different and complex features which a simple neural network cannot learn. Convolutional neural networks are powering core of computer vision that has many applications which include self-driving cars, robotics, and treatments for the visually impaired. The main concept of CNN is to obtain local features from input (usually an image) at higher layers and combine them into more complex features at the lower layers [21] [22].

A typical Convolutional Neural Network architecture consists of the following layers:

a) **Convolutional Layer**

Convolution layer is the core building block of a Convolutional Neural Network which uses convolution operation (represented by *) in place of general matrix multiplication. Its parameters consist of a set of learnable filters also known as kernels. The main task of the convolutional layer is to detect features found within local regions of the input image that are common throughout the dataset and mapping their appearance to a feature map. The convolution operation is given as

$$F(i,j) = (I * K)(i,j) = \sum_m \sum_n I(i+m, j+n) K(m,n) \qquad Eq\ (1)$$

Where *I* is the input matrix (image), K is the 2D filter of size *m x n* and F represents the output 2D feature map. Here the input *I* is convolved with the filter *K* and produces the feature map F. This convolution operation is denoted by I*K.

The output of each convolutional layer is fed to an activation function to introduce non-linearity. There are number of activation functions available but the one which is recognized for deep learning is Rectified Linear Unit (ReLU). ReLU simply computes the activation by thresholding the input at zero. In other words, ReLU outputs 0 if the input is less than 0, and raw output otherwise. It is mathematically given as:

$$f(x) = \max(0, x) \qquad Eq(2)$$

b) **Subsampling (Pooling) Layer**

In CNN, the sequence of convolution layer is followed by an optional pooling or down sampling layer to reduce the spatial size of the input and thus reducing the number of parameters in the network. A pooling layer takes each feature map output from the convolutional layer and down samples it i.e. pooling layer summarizes a region of neurons in the convolution layer. There most common pooling technique is Max Pooling which simply outputs the maximum value in the input region. Other pooling options are average pooling and L2-norm pooling.

c) **Fully Connected Layer**

In fully connected layer each neuron from previous layer is connected to every neuron in the next layer and every value contributes in predicting how strongly a value matches a particular class. The output of last fully connected layer is then forwarded to an activation function which outputs the class scores. Softmax and Support Vector Machines (SVM) are the two main classifiers used in CNN. Softmax function which computes the probability distribution of the n output classes is given as

$$Z^k = \frac{e^{x^k}}{\sum_{i=1}^{n} e^{x^n}} \qquad Eq\ (3)$$

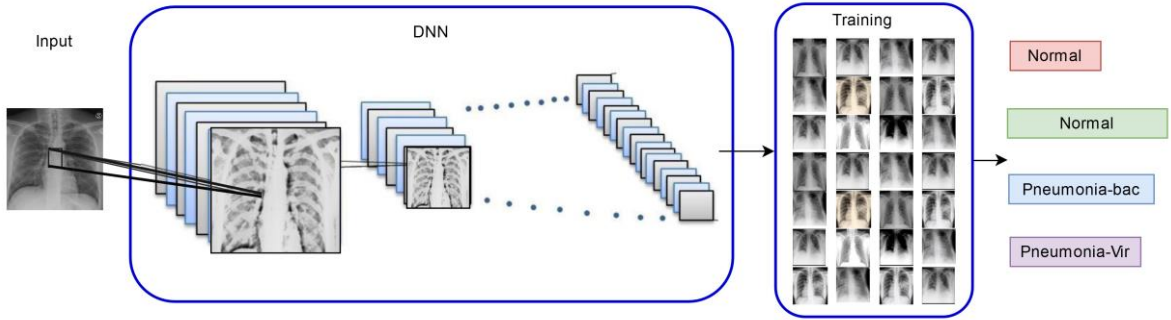

Figure 2: Overview of the proposed methodology

Where x is the input vector and Z is the output vector. The sum of all outputs (Z) equals to 1. The proposed model CoroNet uses Softmax, to predict the class to which the input X-ray image belongs to.

All the layers discussed above are stacked up to make a full CNN architecture. In addition to these main layers mentioned above, CNN may include optional layers like batch normalization layer to improve the training time and dropout layer to address the overfitting issue.

**Model Architecture and Development**

CoroNet is a CNN architecture tailored for detection of COVID-19 infection from chest X-ray images. It is based on Xception CNN architecture [23]. Xception which stands for Extreme version of Inception [24] (its predecessor model) is a 71 layers deep CNN architecture pre-trained on ImageNet dataset. Xception uses depthwise separable convolution layers with residual connections instead of classical convolutions. Depthwise Separable Convolution replaces classic *n x n x k* convolution operation with *1 x 1 x k* point-wise convolution operation followed by channel-wise *n x n* spatial convolution operation. This way the number of operations are reduced by a factor proportional to 1/k.

Residual connections are 'skip connections' which allow gradients to flow through a network directly, without passing through non-linear activation functions and thus avoiding the problem of vanishing gradients. In residual connections, output of a weight layer series is added to the original input and then passed through non-linear activation function as shown in Figure 3.

CoroNet uses Xception as base model with a dropout layer and two fully-connected layers added at the end. CoroNet has 33,969,964 parameters in total out of which 33,969,964 trainable and 54528 are non-trainable parameters. Architecture details, layer-wise parameters and output shape of CoroNet model are shown in Table II. To initialize the model parameters, we used Transfer Learning to overcome the problem of overfitting as the training data was not sufficient.

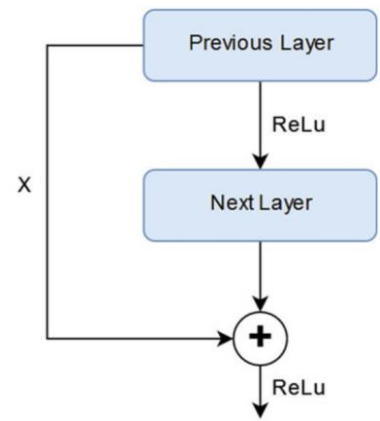

Figure 3: Residual Connection

Table II: Details of CoroNet Architecture

| Layer (type) | Output Shape | Param # |
|---|---|---|
| Xception (Model) | 5 x 5 x 2048 | 20861480 |
| flatten (Flatten) | 51200 | 0 |
| dropout (Dropout) | 51200 | 0 |
| dense (Dense) | 256 | 13107456 |
| dense_1 (Dense) | 4 | 1028 |

Total Parameters: 33,969,964

Trainable Parameters: 33,915,436

Non-trainable Parameters: 54,528

**Implementation and Training**

We implemented three scenarios of the proposed model to detect COVID-19 from chest X-ray images. First model is the main multi-class model (4-class CoroNet) which is trained to classify chest X-ray images into four categories: *COVID-19*, *Normal*, *Pneumonia-bacterial* and *Pneumonia-viral*. The other two models 3-class CoroNet (*COVID-19, Normal* and *Pneumonia*) and binary 2-class CoroNet model (*COVID-19, Normal* and *Pneumonia*) are modifications of the main multi-class model.

The proposed model, CoroNet was implemented in Keras on top of Tensorflow 2.0. The model was pre-trained on ImageNet dataset and then retrained end-to-end on prepared dataset using Adam optimizer with learning rate of 0.0001, batch size of 10 and epoch value of 80. For training, data shuffling was enabled which involves shuffling the data before each epoch. All the experiment and training was done on Google Colaboratory Ubuntu server equipped with Tesla K80 graphics card. We used 4-fold cross-validation [25] approach to assess the performance of our main 4-class model. The training set was randomly divided into 4 equal sets. Three out of four sets were used to train the CNN model while the remaining set was used for validation. This strategy was repeated 4 times by shifting the validation and training sets. Final performance of the model was reported by averaging values obtained from each fold. Plots of accuracy and loss on the training and validation sets over training epochs for Fold 4 are shown in Figure 4.

**RESULTS**

The multi-class classification result of the proposed model was recorded for each Fold and then average numbers were calculated. The performance of the proposed CoroNet on each fold is presented in the form of Confusion matrix (CM) in Figure 5. Overall Accuracy, precision, recall and F-measure computed for each fold by formulae given below are summarized in Table III.

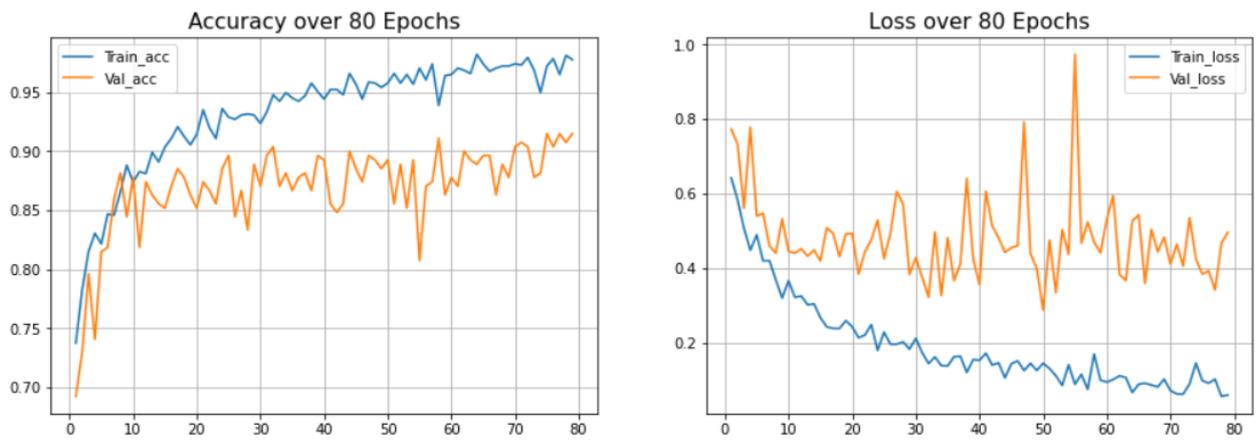

Figure 4: Plots of Accuracy and Loss on Training and validation sets over training epochs for fold 4

$$\text{Accuracy} = \frac{No.\,of\,images\,correctly\,classified}{Total\,no.\,of\,images}$$

$$\text{Precision} = \frac{Sum\,of\,all\,True\,Positives\,(TP)}{Sum\,of\,all\,True\,Positives\,(TP) + All\,False\,Positives\,(FP)}$$

$$\text{Recall} = \frac{Sum\,of\,all\,True\,Positives\,(TP)}{Sum\,of\,all\,True\,Positives\,(TP) + All\,False\,Negatives\,(FN)}$$

$$\text{F-measure} = \frac{2*Precision*Recall}{Precision+Recall}$$

The aforementioned performance metrics are the top metrics used to measure the performance of classification algorithms. The proposed model CoroNet achieved an average accuracy of 89.6%, while as average accuracy, precision, recall and F-measure (F1-Score) for COVID-19 class are 96.6%, 93.17%, 98.25% and 95.6% respectively. The class wise performance of CoroNet is presented in Table IV. The performance for Non COVID-19 pneumonia classes (pneumonia-bacterial and pneumonia-viral) are comparatively lower than other two classes and contributes to the lower overall accuracy. If we combine the pneumonia-bacterial and pneumonia-viral into one single class as Pneumonia class, then the overall accuracy increases significantly. We did slight modification to same 4-class CoroNet model and fine-tuned it for three classes only. In this way we did not have to retrain the model from end-to-end. After fine-tuning, the model was tested on test set comprising of 29 COVID-19, 72 normal and 120 pneumonia cases. The Confusion matrix of 3-class CoroNet is presented in Figure 6(a). After combining the two non-COVID-19 pneumonia infections, the overall accuracy of CoroNet increased from 89.5% to 94.59%. Finally, the confusion matrix for the binary classification problem in detecting COVID-19 positive is presented in Figure 6(b). In addition, precision, recall, F-measure, and accuracy results for all three classification tasks are given in Table V.

Table III: Performance of the CoroNet on each fold

| Folds | Precision (%) | Recall (%) | Specificity (%) | F-measure (%) | Accuracy (%) |
|---|---|---|---|---|---|
| Fold 1 | 88 | 87.7 | 95.7 | 87.6 | 87.3 |
| Fold 2 | 90.8 | 90.7 | 96.7 | 90.7 | 90 |
| Fold 3 | 88.9 | 89 | 96.2 | 88.9 | 89.1 |
| Fold 4 | 92.5 | 92.2 | 97.3 | 92.1 | 92.26 |
| **Average** | **90** | **89.92** | **96.4** | **89.8** | **89.6** |

Table IV: Average class-wise precision, recall, F-measure of 4-class CoroNet

| Class | Precision (%) | Recall (%) | Specificity (%) | F-measure (%) |
|---|---|---|---|---|
| COVID-19 | 93.17 | 98.25 | 97.9 | 95.61 |
| Normal | 95.25 | 93.5 | 98.1 | 94.3 |
| Pneumonia Bacterial | 86.85 | 85.9 | 95 | 86.3 |
| Pneumonia Viral | 84.1 | 82.1 | 94.8 | 83.1 |

Table V: Performance of 4-class, 3-class and binary CoroNet

| Model | Precision (%) | Recall (%) | Specificity (%) | F-measure (%) | Accuracy (%) |
|---|---|---|---|---|---|
| 4-class CoroNet | 90 | 89.92 | 96.4 | 89.8 | 89.6 |
| 3-Class CoroNet | 95 | 96.9 | 97.5 | 95.6 | 95 |
| Binary CoroNet | 98.3 | 99.3 | 98.6 | 98.5 | 99 |

Figure 5: Confusion Matrices of 4-class classification task (a) Fold 1 CM (b) Fold 2 CM (c) Fold 3 CM (d) Fold 4 CM

Figure 6: Confusion matrix results of CoroNet a) 3-class Classification and b) binary classification

**Classification accuracy on Dataset-2**

To check the robustness, we tested our proposed model on another dataset prepared by Ozturk et al. [18]. The dataset-2 contains around 500 normal, 500 pneumonia and 157 COVID-19 chest X-ray images. This dataset contains same COVID-19 X-ray images as in our prepared dataset, however normal and pneumonia X-ray images were collected from ChestX-ray database provided by Wang et al. [26]. After slight modification and fine-tuning, our proposed model achieved an overall accuracy of 90%, The results are illustrated in Table VI and corresponding confusion matrix is given in Figure 7. Table VI shows accuracy, precision, recall and F-measure of CoroNet on Dataset-2.

|            | COVID-19 | Normal | Pneumonia |
|------------|----------|--------|-----------|
| COVID-19   | 33       | 4      | 0         |
| Normal     | 1        | 128    | 21        |
| Pneumonia  | 0        | 7      | 143       |

Figure 7: Confusion matrix result of CoroNet on Dataset-2 [18]

**Table VI: Performance of the CoroNet on Dataset-2 [18]**

| Class              | Precision (%) | Recall (%) | Specificity (%) | F-measure (%) |
|--------------------|---------------|------------|-----------------|---------------|
| COVID-19           | 97            | 89         | 99.6            | 93            |
| Normal             | 92            | 85         | 97.7            | 89            |
| Pneumonia Bacterial| 87            | 95         | 88.7            | 91            |
| Average            | **92**        | **90**     | **95.3**        | **91**        |
| Overall Accuracy   | **90.21%**    |            |                 |               |

**DISCUSSION**

In this study, we proposed a deep model based on Xception architecture to detect COVID-19 cases from chest X-ray images. The proposed model was tested on two datasets and performed exceptionally well on both of them. Our model achieved an accuracy of 89.5%, 94.59% and 99% for 4-classes, 3-classes and binary class classification tasks respectively. Furthermore, our model also achieved an accuracy of 90% on dataset-2. Another positive observation from the results is the precision (PPV) and recall (Sensitivity) for COVID-19 cases. Higher recall value means low false negative (FN) cases and low number of FN is an encouraging result. This is important because minimizing the missed COVID-19 cases as much as possible is the main aim of this research.

The results obtained by our proposed model are superior compared to other studies in the literature. Table VII presents a summary of studies conducted in the automated diagnosis of COVID-19 from chest X-ray images and their comparison with our proposed model CoroNet. Figure 8 shows CoroNet results on some sample images from the test set.

Wang and Wong [10] presented a residual deep architecture called COVID-Net for detection of COVID-19 from chest X-ray images. COVID-Net is one of the early works done on COVID-19 which uses deep neural network to classify chest X-ray images into four categories (COVID, Normal, Pneumonia bacterial and Pneumonia Viral). COVID-Net achieved an accuracy of 83.5% for four classes. Table VIII presents performance comparison of COVID-Net and our proposed model CoroNet on 4-class classification task.

Apostolopoulos and Mpesiana [16] evaluated various state-of-the-art deep architectures on chest X-ray images. With transfer learning their best model VGG19 managed to achieve an accuracy of 93.48% and 98.75% for 3-class and 2-class classification tasks respectively on a dataset consisting of 224 COVID-19, 700 pneumonia and 504 normal X-ray images.

Narin et al. [11] performed same experiment with three different CNN models (ResNet50, InceptionV3, and InceptionResNetV2) and ResNet50 pre-trained on ImageNet database achieved best accuracy of 98% for 2-class classification. Since, they did not include pneumonia cases in their experiment, it is unknown how well their model would distinguish between COVID-19 and other pneumonia cases.

**Table VII: Comparison of the proposed CoroNet with other existing deep learning methods**

| Study | Architecture | Accuracy 3-class (%) | Accuracy 2-class (%) | # Params (in million) |
|---|---|---|---|---|
| Ioannis et al. [43] | VGG19 | 93.48 | 98.75 | 143 |
| Ioannis et al. [43] | Xception | 92.85 | 85.57 | 33 |
| Wang and Wong [42] | Covid-Net (Residual Arch) | NA | 92.4 | 116 |
| Sethy and Behra [45] | ResNet-50 | NA | 95.38 | 36 |
| Hemdan et al. [41] | VGG19 | NA | 90 | 143 |
| Narin et al. [44] | ResNet-50 | NA | 98 | 36 |
| Narin et al. [44] | InceptionV3 | NA | 97 | 26 |
| Ozturk et al [18] | DarkNet | 87.02 | 98.08 | 1.1 |
| **Proposed CoroNet** | CoroNet (Xception) | **89.6** | **99** | 33 |

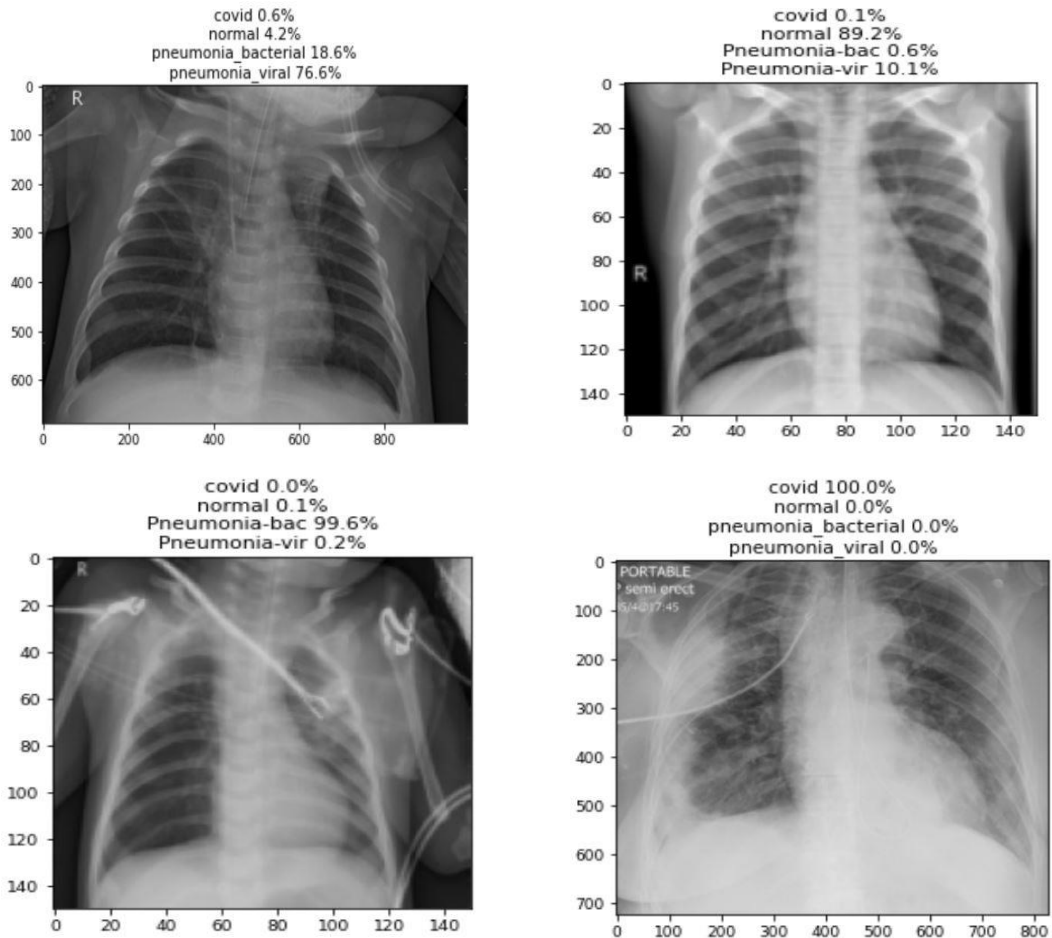

Figure 8: Some images evaluated by CoroNet

CNNs are deep models which perform automatic feature extraction from input data and based on these extracted features, a classifier like Softmax performs classification. Softmax classifier is a default but noncompulsory choice for CNN's and can be replaced by any good classifier like Support Vector Machine (SVM). One such experiment was done by Sethy and Behera [17]. They employed ResNet50 CNN model along with SVM for detection of COVID-19 cases from chest X-ray images. CNN model acts as feature extractor and SVM serves the purpose of classifier. Their model achieved an accuracy of 95.38% on 2-class problem.

Table VIII: 4-class Performance Comparison between Covid-Net and Proposed CoroNet

| Class | COVID-Net [10] | | | CoroNet | | |
|---|---|---|---|---|---|---|
| | Precision (%) | Recall (%) | F-measure (%) | Precision (%) | Recall (%) | F-measure (%) |
| COVID-19 | 80 | 100 | 88.8 | **93.17** | 98.25 | **95.61** |
| Normal | 95.1 | 73.9 | 83.17 | **95.25** | **93.5** | **94.3** |
| Pneumonia Bacterial | 87.1 | 93.1 | 90 | 86.85 | 85.9 | 86.3 |
| Pneumonia Viral | 67.0 | 81.9 | 73.7 | **84.1** | **82.1** | **83.1** |
| # of Parameters | 116 million | | | 33 million | | |
| Accuracy | 83.5% | | | **89.6%** | | |

Ozturk et al [18] proposed a CNN model based on DarkNet architecture to detect and classify COVID-19 cases from X-ray images. Their model achieved binary and 3-class classification accuracy of 98.08% and 87.02%, respectively on a dataset consisting of 125 COVID-19, 500 Pneumonia and 500 normal chest X-ray images.

The promising and encouraging results of deep learning models in detection of COVID-19 from radiography images indicate that deep learning has a greater role to play in fighting this pandemic in near future. Some limitation of this study can be overcome with more in depth analysis which is possible once more patient data (both symptomatic and asymptomatic patients) becomes available.

**CONCLUSION**

As the cases of COVID-19 pandemic are increasing daily, many countries are facing shortage of resources. During this health emergency, it is important that not even a single positive case goes unidentified. With this thing in mind, we proposed a deep learning approach to detect COVID-19 cases from chest radiography images. The proposed method (CoroNet) is a convolutional neural network designed to identify COVID-19 cases using chest X-ray images. The model has been trained and tested on a small dataset of few hundred images prepared by obtaining chest X-ray images of various pneumonia cases and COVID-19 cases from different publically available databases. CoroNet is computationally less expensive and achieved promising results on the prepared dataset. The performance can further be improved once more training data becomes available. Notwithstanding the encouraging results, CoroNet still needs clinical study and testing but with higher accuracy and sensitivity for COVID-19 cases, CoroNet can still be beneficial for radiologists and health experts to gain deeper understandings into critical aspects associated with COVID-19 cases.

**Source Code and Dataset:** For further research in this area we have made the source code, trained model and dataset available at https://github.com/drkhan107/CoroNet

**Funding Sources:** This research did not receive any specific grant from funding agencies in the public, commercial, or not-for-profit sectors.

**Declaration of Competing Interest**

The authors have no conflict of interest to disclose.

**References**

1. WHO updates on COVID-19  https://www.who.int/emergencies/diseases/novel-coronavirus-2019/events-as-they-happen (Last Accessed: 3 Apr 2020)

2. Mahase, Elisabeth. "Coronavirus: covid-19 has killed more people than SARS and MERS combined, despite lower case fatality rate." (2020). doi: https://doi.org/10.1136/bmj.m641.

3. COVID-19 symptoms  https://www.who.int/health-topics/coronavirus#tab=tab_3   Last Accessed: 3 Apr 2020

4. Wang W, Xu Y, Gao R, Lu R, Han K, Wu G, Tan W. Detection of SARS-CoV-2 in different types of clinical specimens. Jama. 2020 Mar 11.

5. Corman VM, Landt O, Kaiser M, Molenkamp R, Meijer A, Chu DK, Bleicker T, Brünink S, Schneider J, Schmidt ML, Mulders DG. Detection of 2019 novel coronavirus (2019-nCoV) by real-time RT-PCR. Eurosurveillance. 2020 Jan 23;25(3):2000045.

6. Bernheim A, Mei X, Huang M, Yang Y, Fayad ZA, Zhang N, Diao K, Lin B, Zhu X, Li K, Li S. Chest CT findings in coronavirus disease-19 (COVID-19): relationship to duration of infection. Radiology. 2020 Feb 20:200463.

7. Xie X, Zhong Z, Zhao W, Zheng C, Wang F, Liu J. Chest CT for typical 2019-nCoV pneumonia: relationship to negative RT-PCR testing. Radiology. 2020 Feb 12:200343.

8. Fang Y, Zhang H, Xie J, Lin M, Ying L, Pang P, Ji W. Sensitivity of Chest CT for COVID-19: Comparison to RT-PCR. Radiology 2020 Feb. 19:200432. [Epub ahead of print], doi: https://pubs.rsna.org/doi/full/10.1148/radiol.2020200432.

9. XieX, Zhong Z, Zhao W, Zheng C, Wang F, Liu, J. Chest CT for typical 2019-nCoV pneumonia: relationship to negative RT-PCR testing. Radiology 2020 Feb. 7 (Epub ahead of print), doi: https://pubs.rsna.org/doi/10.1148/radiol.2020200343

10. Wang L, Wong A. COVID-Net: A Tailored Deep Convolutional Neural Network Design for Detection of COVID-19 Cases from Chest Radiography Images. arXiv preprint arXiv:2003.09871. 2020 Mar 22.

11. Narin A, Kaya C, Pamuk Z. Automatic Detection of Coronavirus Disease (COVID-19) Using X-ray Images and Deep Convolutional Neural Networks. arXiv preprint arXiv:2003.10849. 2020 Mar 24.

12. Shen D, Wu G, Suk HI. Deep learning in medical image analysis. Annual review of biomedical engineering. 2017 Jun 21;19:221-48.

13. Rajpurkar P, Irvin J, Zhu K, Yang B, Mehta H, Duan T, Ding D, Bagul A, Langlotz C, Shpanskaya K, Lungren MP. Chexnet: Radiologist-level pneumonia detection on chest X-rays with deep learning. arXiv preprint arXiv:1711.05225. 2017 Nov 14.

14. Wang H, Xia Y. Chestnet: A deep neural network for classification of thoracic diseases on chest radiography. arXiv preprint arXiv:1807.03058. 2018 Jul 9.

15. Hemdan EE, Shouman MA, Karar ME. Covidx-net: A framework of deep learning classifiers to diagnose covid-19 in X-ray images. arXiv preprint arXiv:2003.11055. 2020 Mar 24.

16. Apostolopoulos ID, Mpesiana TA. Covid-19: automatic detection from X-ray images utilizing transfer learning with convolutional neural networks. Physical and Engineering Sciences in Medicine. 2020 Apr 6:1

17. Sethy PK, Behera SK. Detection of coronavirus disease (covid-19) based on deep features. Preprints. 2020 Mar 19;2020030300:2020.

18. Ozturk T, Talo M, Yildirim EA, Baloglu UB, Yildirim O, Acharya UR. Automated detection of COVID-19 cases using deep neural networks with X-ray images. Computers in Biology and Medicine. 2020 Apr 28:103792

19. Joseph Paul Cohen, Paul Morrison and Lan Dao COVID-19 image data collection, arXiv:2003.11597, 2020 https://github.com/ieee8023/covid-chestxray-dataset.

20. Chest X-ray images (pneumonia). https://www.kaggle.com/paultimothymooney/chest-xray-pneumonia. Last Accessed: 1 Apr 2020

21. Wan J, Wang D, Hoi SC, Wu P, Zhu J, Zhang Y, Li J. Deep learning for content-based image retrieval: A comprehensive study. InProceedings of the 22nd ACM international conference on Multimedia 2014 Nov 3 (pp. 157-166).

22. Wani MA, Bhat FA, Afzal S, Khan AI. Advances in Deep Learning. Springer; 2020.

23. Chollet F. Xception: Deep learning with depthwise separable convolutions. InProceedings of the IEEE conference on computer vision and pattern recognition 2017 (pp. 1251-1258).

24. Szegedy C, Vanhoucke V, Ioffe S, Shlens J, Wojna Z. Rethinking the inception architecture for computer vision. InProceedings of the IEEE conference on computer vision and pattern recognition 2016 (pp. 2818-2826).

25. Duda, R. O., Hart, P. E., Stork, D. G., 2001. Pattern classification 2nd edition, New York, John Wiley and Sons

26. Wang X, Peng Y, Lu L, Lu Z, Bagheri M, Summers RM. Chestx-ray8: Hospital-scale chest X-ray database and benchmarks on weakly-supervised classification and localization of common thorax diseases. InProceedings of the IEEE conference on computer vision and pattern recognition 2017 (pp. 2097-2106).